\documentclass[journal]{IEEEtran}
\usepackage{cite}

\ifCLASSINFOpdf
  \usepackage[pdftex]{graphicx}
\else
  \usepackage[dvips]{graphicx}
\fi
\usepackage{amsmath}

\usepackage{xcolor}

\newcommand{\rev}[1]{\textcolor{black}{#1}}

\begin{document}
%
\title{Beyond white- and black-box modeling tools in optical communications and optical computing: physics-informed data-driven modeling} 
%
%
%
\author{Isidora Teofilovic, Sergio Hernandez Fernandez, Metodi P. Yankov, Christophe Peucheret~\IEEEmembership{Member,~IEEE},  Darko Zibar, Francesco Da Ros~\IEEEmembership{Senior Member,~IEEE},
\thanks{This work was supported by VILLUM FONDEN with grants VI-POPCOM (VIL54486), and YIP OPTIC-AI (VIL29334), by Horizon Europe research and innovation project PROMETHEUS (grant n. 101070195), the Swedish Research Council, BRAIN grant no. 2022-04798, and the MSCA Doctoral network MINDnet (grant n. 101226674).}%
\thanks{F. Da Ros, I. Teofilovic, and D. Zibar are with the Department of Electrical and Photonics Engineering, Technical University of Denmark, 2800 Kgs. Lyngby, Denmark. ({fdro, isteo, dazi}@dtu.dk)}
\thanks{S. Hernandez Fernandez was with the Department of Electrical and Photonics Engineering, Technical University of Denmark. He is now with Cisco Denmark.}
\thanks{M.P. Yankov was with the Department of Electrical and Photonics Engineering, Technical University of Denmark. He is now with Ciena Denmark.}
\thanks{C. Peucheret is with the CNRS, FOTON-UMR6082, Univ. Rennes, 22305 Lannion, France}
\thanks{Manuscript received April 19, 2005; revised August 26, 2015.}}

\markboth{Journal of \LaTeX\ Class Files,~Vol.~14, No.~8, August~2015}%
{Shell \MakeLowercase{\textit{et al.}}: Bare Demo of IEEEtran.cls for IEEE Journals}
%



\maketitle

\begin{abstract}
Efficient optimization and control of photonic computing and communication systems increasingly rely on accurate surrogate models/digital twins. While data-driven models may achieve faster inference than traditional physics-based methods, they typically suffer from poor training data efficiency and limited generalizability. To address this trade-off, physics-informed data-driven modeling has emerged as a powerful hybrid paradigm. This paper presents a comparative analysis of these three modeling paradigms across three \rev{test} use cases: optical amplifiers, directly modulated lasers, and interferometer meshes. By evaluating model complexity, data efficiency, generalizability, and modularity, this work provides a detailed analysis of the respective trade-offs and highlights the advantages of combining physical insight with data-driven learning.
\end{abstract}


%
\IEEEpeerreviewmaketitle

\section{Introduction}

\IEEEPARstart{{P}}{rogramming}, controlling, and optimizing photonic integrated circuits (PICs) and photonic systems, e.g. for communication and computing, are core operations whose complexity grows together with advances in PIC and system design. The number of components integrated on a single PIC has already crossed into the regime of very large-scale integration, i.e., beyond 10,000~\cite{shekhar2024roadmapping}. Simultaneously, optical communication systems are integrating an expanding set of multiplexing techniques - spanning space-division and multi-band approaches - which increases both the number and the complexity of the components required ~\cite{agrell2024roadmap,puttnam2021space}. These needs, coupled with the growing performance requirements, have driven a strong research focus towards building offline (in silico) models - so-called digital twins - of components, subsystems, and full systems that can be used for optimization and control~\cite{borraccini2024local,wang2024digital,faruk2023measurement,fan2020advancing, MoralisPegios2022,Lee2024,Bandyopadhyay2021}. Alternatively, online (in-situ) programming and optimization methods that do not rely on offline models~\cite{qiu2025trustopt,Milanizadeh2020,demoura2022OECC,Fyrillas2024} are still highly sought after. However, iterative methods suffer from a relatively long convergence time~\cite{demoura2022OECC, jovanovic2021gradient}. Online methods are still extremely effective as a fine-tuning step once a `good' initial condition is provided, e.g., through an offline model~\cite{demoura2022OECC,Soltani2021OL,yankov2025OFT}. Therefore, accurate, fast, generalizable and differentiable models are strongly needed.

The renewed interest in modeling has also been driven by the widespread use of machine learning (ML) tools, which have extended the capabilities of building data-driven models. The capacity of ML models to learn complex mappings and nonlinear dynamics directly from data enables the creation of physics-agnostic surrogate models.  These data-driven models have demonstrated several key advantages over physical models: they enable accurate modeling where physical models are lacking (e.g., Bi-doped fiber amplifiers~\cite{Donodin2023JEOSRP}), offer computational efficiency for systems with complex physical dynamics (e.g., Raman amplifiers~\cite{Zibar2019}), support inverse-system modeling directly from measurements~\cite{Zibar2019}, and are inherently differentiable - unlike most physical models (e.g., laser rate equations~\cite{coldren1997diode}) - thereby unlocking gradient-based, end-to-end optimization. Nevertheless, purely data-driven models, commonly referred to as black-box (BB) models, as opposed to purely physics-based white-box (WB) models, are limited by their need to learn the underlying physics directly from data. This requirement results in relatively high model complexity, large training datasets that generally scale unfavorably with system size, and poor generalization, as we will discuss in the following. Consequently, a modeling paradigm that integrates physical knowledge with data-driven learning has attracted considerable interest in recent years~\cite{raissi_physics-informed_2019}. They are commonly referred to as grey-box (GB) models, i.e. a combination of BB and WB models.

\begin{figure*}[!tb]
    \centering
    \includegraphics[width=0.97\linewidth]{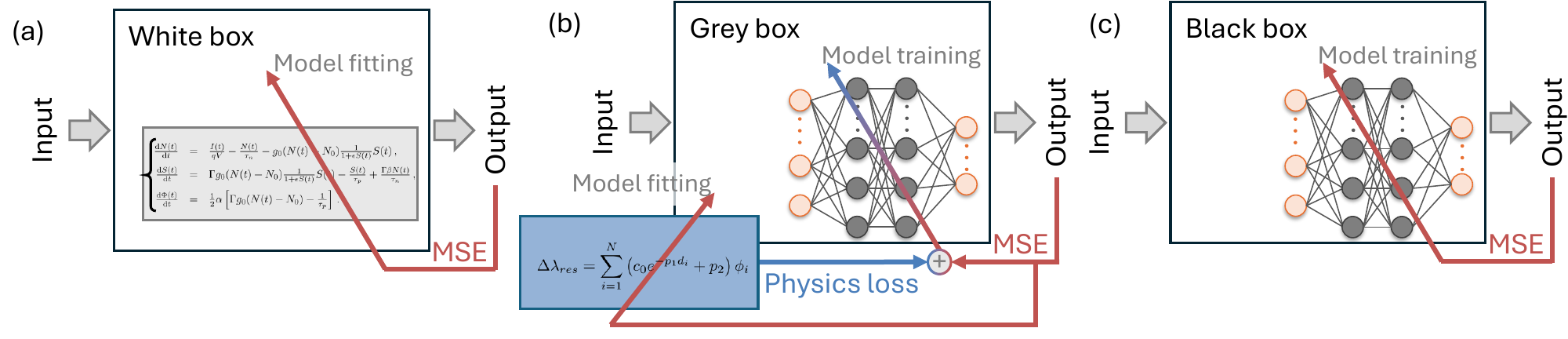}
    \caption{Main category of modeling approaches: (a) physical models (WB) where a system of physical equations is fitted to data; (b) physics-informed data-driven (GB) model where a ML-based model is trained by data and underlying physics, which can also be simultaneously fitted to data; (c) data-driven (BB) model where a ML-based model is trained on data based on the mean square error (MSE) between target and predictions. }
    \label{fig:modelingapproaches}
\end{figure*}
Here, we review the three modeling paradigms - physics-based (WB), data-driven (BB), and physics-informed data-driven - qualitatively assessing the general strengths and weaknesses of each paradigm, and presenting quantitative results across three use cases: optical amplifiers, directly modulated laser links, and interferometer meshes. This work reviews, contextualizes, and expands on our previous contributions on the topic. It also extends our conference submission~\cite{fdroOFC} through a systematic cross-case comparison and a deeper investigation of physics-informed data-driven models, especially for interferometer meshes, with particular focus on their generalizability and modularity. The paper is structured as follows: Section~\ref{sec:Models} introduces the three modeling paradigms focusing on the three use cases for each; Section~\ref{sec:Challenges} systematically discusses challenges and proposed solutions with a strong focus on the benefits of combining physical knowledge with data-driven learning; Section~\ref{sec:Conclusions} summarizes the discussion and draws the main conclusions.

\section{Modeling paradigms}
\label{sec:Models}

The three main modeling paradigms applied to photonic systems are sketched in Fig.~\ref{fig:modelingapproaches}. They are classified by the amount of physical knowledge they include. The first and historically most common class of models relies purely on the physics (Fig.~\ref{fig:modelingapproaches}(a)). These white-box (WB) models rely on physics-based equations whose parameters can be fitted to measurements~\cite{deMoura2023JLT,marchisio2025comprehensive}. Alternatively, over the last couple of decades, ML models that learn input-output mappings directly from data have become popular for surrogate modeling~\cite{borraccini2024local,wang2024digital}. They generally rely on different flavors of neural network (NN) architectures that are trained directly from input/output data and are physics-agnostic (Fig.~\ref{fig:modelingapproaches}(c)).  Finally, a more recent paradigm relies on the concept of physics-informed data-driven learning (Fig.~\ref{fig:modelingapproaches}(b)). Physics-informed models, referred to as grey-box (GB) models, aim at combining preexisting physical knowledge, even partial, with direct learning from the data. Therefore, they effectively lie at the intersection of WB and BB models. In the following subsections, we will discuss the strengths and weaknesses of each modeling paradigm by focusing on three specific modeling use cases: optical amplifiers, and more specifically Raman amplifiers; directly-modulated lasers as transmitters for short-link communication; and interferometer meshes as matrix-vector multipliers for optical computing. The main metrics that can be of interest in the perspective of controlling or optimizing  photonic systems using different flavors of offline models are: (1) model accuracy; (2) data efficiency; (3) computational efficiency; (4) model generalizability; and (5) model differentiability. Model accuracy can be evaluated by comparing the model prediction with the ground truth provided by measurements. The main metric considered is the root mean squared error (RMSE), or its normalized version, normalized RMSE (NRMSE).  Data- and computational-efficiency relate to the amount of data and computational resources (e.g. computing time and memory) required to fit the model to the data. Computational efficiency can also be considered in the context of applying the model after the training process, e.g. at the inference stage. This latter aspect is particularly relevant for iterative optimization processes where inference is repeatedly performed. Model generalizability describes the ability of the model to generalize beyond the specific realization of the photonic system whose data it was trained on. This property is particularly relevant in the context of BB models (as will be discussed more in the following and in Section \ref{sec:Challenges}) as models that are overfitted to a single device and its operating conditions, require an unfavorable scaling of the number of models that need to be trained/fitted. Finally, differentiability is directly related to ease of optimization. Differentiability allows the application of powerful gradient-based algorithms to optimize through the model without resorting to gradient-free optimizers which generally increase the computing load and may degrade accuracy~\cite{jovanovic2021gradient}. These metrics will be qualitatively discussed at a high-level for each modeling paradigm and quantitative results will be presented for the three use cases considered here.

\subsection{White-box models}
\label{subsec:WB}

WB models rely on a well-defined set of equations, which describe the physics of the system. The physical equations are parametrized by physical quantities that can sometimes be recovered by fitting the model to data (see Fig.~\ref{fig:modelingapproaches}(a)). As the photonic systems grow in complexity,  WB models generally become either less accurate or more involved. The latter option directly translates into additional computational resources for the numerical solvers, as well as potentially larger datasets for fitting an increasing number of physical parameters. Additionally, WB models are generally challenging to define in a differentiable format since partial differential or integral equations may not ensure differentiability. While current progress in automatic differentiation is enabling implementation of numerical solvers for differential equations that can provide gradient estimation, e.g. through the framework of PyTorch~\cite{paszke_pytorch_2019} or JAX~\cite{bradbury_jax_2018}, accuracy is not guaranteed by the numerical algorithms. However, WB models, by utilizing the underlying physics that generalizes across different realizations of the same physical system, allow for models that can be easily translated across devices, at most with a fast re-fitting of few parameters.

\subsubsection{Optical amplifiers}

\begin{figure}[!th]
    \centering
    \includegraphics[width=0.8\linewidth]{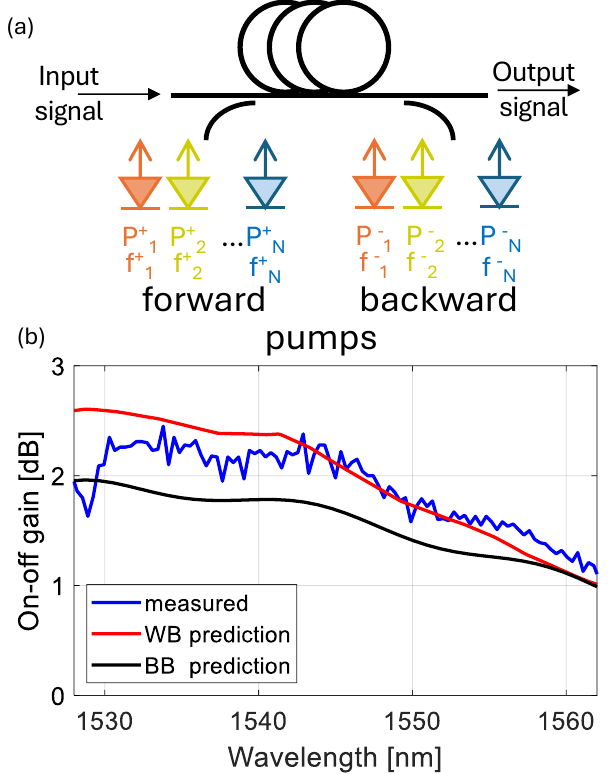}
    \caption{Raman amplifier - (a) Raman amplifier setup relying on multi-pump ($P_i$,$f_i$). Both forward and backward pumping are highlighted. (b) Example of Raman on-off gain from a backward-pumped 50~km standard single-mode fiber: measurements are compared with fitted WB model and learned BB model (for details on the amplifier refer to~\cite{deMoura2023JLT})}
    \label{fig:RamanAmpl}
\end{figure}

The first use case we consider is the modeling of optical amplifiers, and more specifically Raman amplifiers. Raman amplifiers make use of stimulated Raman scattering taking place in optical fibers to transfer energy from an optical pump to the signals to be amplified. Pumps can be co-propagating with the signal (forward-pumping) or counter-propagating (backward pumping) or both (bi-directional pumping). A typical setup of a Raman amplifier is shown in Fig.~\ref{fig:RamanAmpl}(a), highlighting the key sets of parameters that need to be optimized for achieving desired performance: pump power levels and pump frequencies.

Accurate WB models of Raman amplifiers are well established~\cite{zirngibl1998analytical,bromage2004analytical} and rely on solving systems of coupled partial differential equations describing the signal power ($P_s$), pump power ($P_p$) and amplified spontaneous emission power ($P_A$) evolution along the fiber, such as the one provided in (\ref{eq:Raman}) in the restrictive case of single wavelength pumping. $P^+$ and $P^{-}$ refer to co- and counter-propagating waves with respect to the signal for both pumps and noise, respectively. The system in (\ref{eq:Raman}) can then be expanded to include multiple pump and signal wavelengths by increasing the number of equations.

\begin{equation}
\label{eq:Raman}
\begin{array}{rcl}
\displaystyle \frac{\mathrm{d} P_s}{\mathrm{d}z} &=& -\alpha_s P_s + C_R(\lambda_s,\lambda_p)\left[P_p^{+}+P_p^{-}\right]P_s\,, \\[8pt]

\displaystyle \pm \frac{\mathrm{d} P_p^{\pm}}{\mathrm{d}z} &=& -\alpha_p P_p^{\pm}
- \left(\frac{\lambda_s}{\lambda_p}\right) C_R(\lambda_s,\lambda_p) P_s P_p^{\pm}\,,\\[8pt]

\displaystyle \pm \frac{\mathrm{d} P_A^{\pm}}{\mathrm{d}z} &=& -\alpha_A P_A^{\pm}
+ C_R(\lambda_A,\lambda_p) P_p P_A^{\pm} \\

&& + 2h\nu_A C_R(\lambda_A,\lambda_p)[1+\eta(T)] B_{\mathrm{ref}} P_p\,.
\end{array}
\end{equation}

The equations in (\ref{eq:Raman}) can be fitted to measurement data, e.g. by fitting the fiber attenuation $\alpha$ and the Raman efficiency $C_R$. In Fig.~\ref{fig:RamanAmpl}(b), a measured on-off gain is compared  with the fitted WB model and a learned BB model (see Section~\ref{subsec:BB}). The measurement conditions and model detail can be found in~\cite{deMoura2023JLT}. However, solving this WB model is generally computationally intensive, especially as the number of pumps increases and if backward pumping is considered. The latter case transforms the system of equations into a boundary value problem that may require using e.g., a shooting algorithm~\cite{liu2003effective}, further increasing the computational complexity. Finally, the Raman efficiency is generally defined either as an integral equation or through experimental measurements (e.g., a lookup table), making the system non-differentiable with respect to the pumping wavelengths. While this work will focus primarily on Raman amplifiers, similar considerations can be applied to WB modeling of other optical amplifier technologies, e.g. erbium-doped fiber amplifiers (EDFAs)~\cite{saleh1990analytical} or semiconductor optical amplifiers (SOAs)~\cite{cassioli2000time}.

For Raman amplifiers the modeling goal is to predict the gain spectrum of the amplifier as a function of the input power spectral density (PSD) and the pumping conditions, e.g. pump powers and wavelengths. The prediction accuracy is evaluated in terms of the NRMSE between the predicted on-off gain and the measured one (ground truth).

\subsubsection{Directly-modulated lasers}

As a second use case, we consider the modeling of directly-modulated lasers (DMLs). DMLs are particularly promising for intensity-modulation direct-detection (IM/DD) links where a compact footprint and potentially lower power consumption compared to external modulation may be beneficial. However, the complex dynamics of DMLs introduce significant linear (e.g., through electro-optic bandwidth limitation) and nonlinear (e.g., through intrinsic laser dynamics) signal distortion.

\begin{figure}
    \centering
    \includegraphics[width=0.97\linewidth]{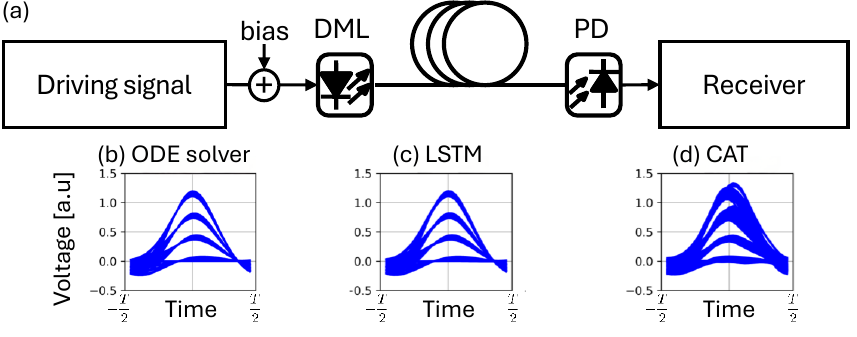}
    \caption{DML - (a) Setup of short-reach link based on DMLs, highlighting the key inputs to a DML: driving signal and bias current. Eye diagrams from the WB DML model (ODE solver, (b)) compared to BB models (long short-term memory NN (c) and convolutional-attention transformer (d)). For details refer to~\cite{fernandez2024PTL}.}
    \label{fig:DML}
\end{figure}

In the context of maximizing the throughput of DML-based links (Fig.~\ref{fig:DML}(a)), e.g. by optimizing transmitter and receiver, accurate and computationally efficient DML models are highly sought after.
WB models of DMLs based on rate equations have been intensively refined~\cite{Petermann1988,Nagarajan1991}, e.g., leading to the system of equations in (\ref{eq:DML}) describing the temporal evolution of carrier density $N(t)$, photon density $S(t)$, and optical phase $\Phi(t)$, see~\cite{coldren1997diode} for full description. 

\begin{equation}
\label{eq:DML}
\begin{array}{rcl}
\frac{\mathrm{d}N(t)}{\mathrm{d}t}&=& \frac{I(t)}{qV} - \frac{N(t)}{\tau_n} - g_0 (N(t) - N_0)\frac{1}{1 + \epsilon S(t)}S(t)\,,\\[8pt]
\frac{\mathrm{d}S(t)}{\mathrm{d}t} &=& \Gamma g_0 (N(t) - N_0)\frac{1}{1 + \epsilon S(t)}S(t) - \frac{S(t)}{\tau_p} + \frac{\Gamma \beta N(t)}{\tau_n}\,, \\[8pt]
\frac{\mathrm{d}\Phi(t)}{\mathrm{d}t}&=& \frac{1}{2}\alpha \left[ \Gamma g_0 (N(t) - N_0) - \frac{1}{\tau_p} \right]\,.\\
\end{array}
\end{equation}

As for the Raman system, the gain parameter $g$, the linewidth enhancement factor $\alpha$, the carrier lifetime $\tau_n$, etc. can be extracted from the measurements~\cite{cartledge1997extraction,marchisio2023particle}. However, measuring the main quantities in (\ref{eq:DML}), e.g. $N(t)$, $S(t)$ and $\Phi(t)$ is rather involved. Furthermore, numerically solving the system of equations is also computationally demanding, requiring a numerical ordinary differential equation (ODE) solver. Finally, (\ref{eq:DML}) does not allow a simple differentiable implementation, especially if the goal is to optimize the driving current $I(t)$.

In the following discussion, we will refer to the modeling of the distributed feedback laser considered in~\cite{fernandez2024PTL}, where the numerical solution of (\ref{eq:DML}) is considered the ground truth. The model accuracy will then be related to the normalized mean squared error (NMSE) between the predicted output optical field given the electrical driving signal (including biasing term). A visual example is shown in Fig.~\ref{fig:DML}(b)-(d) where 4-pulse-amplitude modulated (PAM) eye diagrams are shown comparing the WB model (ODE solver) with two BB models (see further discussion below).

\subsubsection{Interferometer meshes}
Meshes of optical interferometers are a key component in various photonic applications, from optical matrix-vector multipliers as machine-learning accelerators~\cite{brunner2025roadmap, clements_optimal_2016, shokraneh_diamond_2020, feng_compact_2022, mojaver_addressing_2023, wu_real-valued_2023, fldzhyan_low-depth_2024, marchesin_braided_2025}, to quantum unitary transformations~\cite{kumar2021QuiX} and multiple-input multiple-output mode decomposition~\cite{annoni2017unscrambling}.

\begin{figure}
    \centering
    \includegraphics[width=0.97\linewidth]{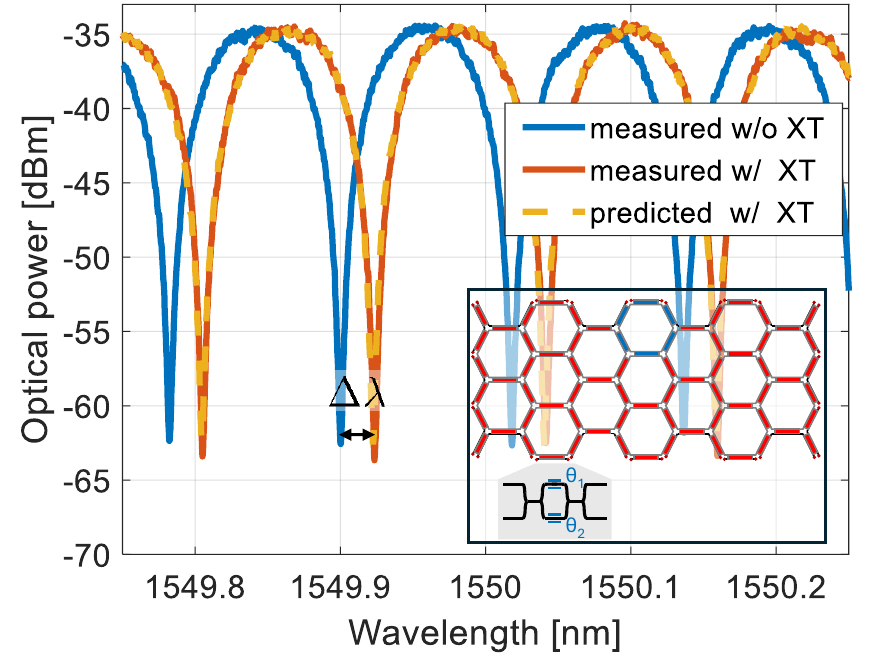}
    \caption{Hexagonal MZI mesh - Transfer function of the emulated micro-ring resonator from one hexagonal cell: measurements without and with crosstalk, as well as predicted crosstalk from WB model are compared. Inset: Hexagonal mesh of MZI. Red and blue rectangles represent MZIs with two degrees of freedom $\theta_1$ and $\theta_2$. Blue MZIs indicate the emulated MRR, red MZIs the crosstalk interferometers.}
    \label{fig:MZImesh}
\end{figure}

The most common realizations rely on Mach-Zehnder interferometers (MZIs) using thermo-optic phase shifters, as shown in Figure~\ref{fig:MZImesh}. Each MZI is normally implemented with two degrees of freedom (see inset in Figure~\ref{fig:MZImesh}), to realize complex transformations.
The transfer function of a single MZI is well defined, e.g. as in (\ref{eq:MZI}) where $\theta_1$ and $\theta_2$ describe the phase applied by the two phase shifters and $x_i$ and $y_i$ represent the input and output ports of the MZI, respectively. 
\begin{equation}
\begin{bmatrix} y_1 \\ y_2 \end{bmatrix}
= \frac{1}{2}
\begin{bmatrix} 1 & \mathrm{j} \\ \mathrm{j} & 1 \end{bmatrix}
\begin{bmatrix} \mathrm{e}^{\mathrm{j}\theta_1} & 0 \\ 0 & \mathrm{e}^{\mathrm{j}\theta_2} \end{bmatrix}
\begin{bmatrix} 1 & \mathrm{j} \\ \mathrm{j} & 1 \end{bmatrix}
\begin{bmatrix} x_1 \\ x_2 \end{bmatrix}\,.
\label{eq:MZI}
\end{equation}

A simple WB model of interferometer meshes can be implemented considering the cascade of MZI transfer functions realized as the product of the single transfer functions. Such a model is rather simple and therefore computationally efficient. However, as the size of the matrix to implement scales up and MZIs are packed closely together, crosstalk (XT) effects quickly become non-negligible~\cite{perez_multipurpose_2017,Ali_JLT}. Thermal XT due to thermal diffusion, electrical XT due to the dense voltage delivery network, and optical XT due to waveguide crossing make simple WB models inaccurate even for small meshes, e.g. the $3 \times 3$ mesh in \cite{Ali_JLT}.

Thermal effects can be included through coupling the optical simulations with a thermal diffusion model. However, the simple analytical expression in (\ref{eq:MZI}) needs to be replaced by numerically solving the field propagation through finite-difference time-domain simulations~\cite{marchisio2025comprehensive}.

Alternatively, thermal XT can be independently fitted considering the thermal-decay model (ThDM) proposed in\cite{Ali_JLT}, where only the exponential decay of the thermal diffusion equation is considered. The choice is confirmed by the extensive numerical thermo-optic simulation in~\cite{marchisio2025comprehensive}, which indeed shows the expected exponential decay with distance. Equation (\ref{eq:ThDM_temp}) shows the evolution of the temperature $T$ with distance $d$ from the heating element at temperature $T_0$, and $p_i$ represent the fitting parameters.

\begin{equation}
T(d) = (p_1 e^{-p_2 d} + p_3) \cdot T_0\,.
\label{eq:ThDM_temp}
\end{equation}

Since the phase is directly proportional to the temperature, such a model can be directly applied to relate the impact of the phase applied to an XT interferometer (XT-MZI), on the phase of the MZI under test. Once the thermal XT is fitted, correction terms can be added to the WB model~\cite{Ali_JLT}, thus moving away from a purely WB model and towards a GB model (see Section ~\ref{subsec:GB}).
In the following, we consider two interferometer-mesh designs: a $3 \times 3$ rectangular mesh composed of 9 MZIs~\cite{Ali_JLT}, and a hexagonal mesh composed of 72 MZIs (`Smartlight', from iPronics~\cite{perez_multipurpose_2017}). In the case of the $3 \times 3$ mesh, no specific design optimization has been performed to address crosstalk.  The models are therefore trained to capture the full input-output mapping of the mesh as a function of the applied voltages. In the case of the hexagonal mesh, instead, its design has been optimized to minimize thermal XT, e.g., by adding thermal isolation with trenches, and the voltage-to-phase mapping has been pre-calibrated. However, even for such an optimized design, phase-sensitive applications, such as the emulation of a micro-ring resonator (MRR), are affected by crosstalk from tuning other interferometers on the same PIC~\cite{teofilovic2024JLT}. The hexagonal mesh is shown in Fig.~\ref{fig:MZImesh} (inset), with the MRR emulated through the hexagonal cell marked in blue (6 MZIs) and the XT-MZIs (remaining 66 MZIs), marked in red. The resonance wavelength of the emulated MRR is shifted by the thermal XT and the error in predicting the shift in resonance wavelength ($\Delta\lambda$) will be the chosen metric for comparing the three modeling paradigms. Visual examples of transfer functions are shown in Fig.~\ref{fig:MZImesh}, comparing the cases without XT, with XT and the response without XT shifted by the predicted $\Delta\lambda$ provided by the ThDM (WB).

\begin{figure*}[!htb]
    \centering
    \includegraphics[width=0.97\linewidth]{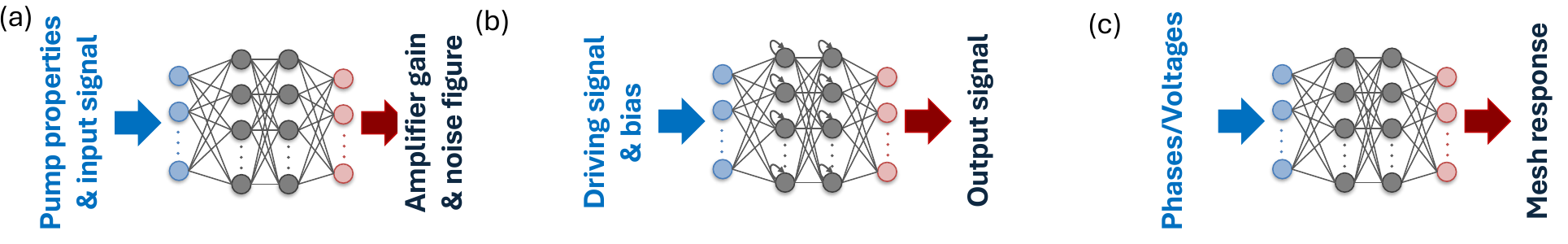}
    \caption{Direct neural network models mapping inputs and/or parameters into the predicted output: (a) Raman amplifier; (b) DML-based link; (c) XT in an MZI mesh.}
    \label{fig:NNsketch}
\end{figure*}

\subsection{Black-box models}
\label{subsec:BB}
With the increased availability of computing power and the broad access to frameworks for machine-learning tools, data-driven BB models have seen a growing interest. Training a NN to learn the input-output mapping of a physical system can provide an accurate, physics-agnostic model of the system. The physics-agnostic property, however, comes at the expense of requiring a large training dataset of measurement data, and dedicated computing resources. Furthermore, generalizability of the models is a significant challenge as BB models are prone to overfit to the specific realization of the physical system. Nevertheless, the key advantages of these models are their computing efficiency at the inference stage, i.e., after having been trained, and their inherent differentiable formulation, which allows for easy optimization using gradient-based methods.
In order to build accurate models, a representative and comprehensive training dataset is critical. Given an input vector $\mathbf{X}$ to be mapped into an output $\mathbf{Y}$ through the learned BB model $\mathbf{Y} = \tilde{f}(\mathbf{X})$, the BB model $\tilde{f}$ accurately approximates input-output mapping $f$ to be learned if the training dataset is representative of the full hyper-space defined by all the input's dimensions~\cite{Zibar2019}. This requirement automatically hints at a poor scaling of the training dataset, i.e., a poor training data efficiency, with the number of input dimensions and with the complexity of $f$. Note that, regardless of the size of the training dataset, BB models are not proven to ensure physical consistency, i.e. the prediction may not follow physical laws.

In the following, we will provide some perspective on BB modeling applied to the three use cases already introduced  in Section~\ref{subsec:WB}, as well as briefly reviewing the state of the art for each case.

\subsubsection{Optical amplifiers}
Optical amplifiers are one of the first photonic systems modeled through NN, with early demonstrations dating back to 2006 for Raman~\cite{Zhou2006} and SOAs\cite{Ababneh2006}. BB NN-based models have been applied to learn both the direct and inverse model of amplifiers, e.g. learning the mapping from amplifier parameters and input signal PSD onto output/gain response (direct, shown in Fig.~\ref{fig:NNsketch}(a) and learning to predict the required amplifier parameters/input PSD for a given target response (inverse, not shown).
More specifically, extensive work has been reported on NN-based models of Raman amplifiers~\cite{Zhou2006,Chen2018,Zibar2019,deMoura2020JLT,deMoura2021OL,Soltani2021OL,deMoura2023JLT,wu2025inverse}, SOAs~\cite{Ababneh2006,zhao2023,Saghiran2025OE,katsura2024CLEO}, EDFAs~\cite{You2018ECOC,ionescu2020CLEO,DaRos2020ECOC,zhu2020,Yankov2021JLT,yu2021JOCN,Rabbani2025OE}, but also hybrid amplifiers~\cite{Ionescu2019SubOptics,DaRos2021OFC,minakhmetov2023OFC,Sarma2025ApplOpt}. More recently the focus has also been extended towards newer amplification technologies, e.g. thulium-\cite{Radovic2024ONDM}, praseodymium-\cite{Sarma2025ApplOpt} and bismuth-doped\cite{Donodin2023JEOSRP} fiber amplifiers (DFAs), as well as parametric amplifiers\cite{Tay2022Optik,sui2022Optik}. While for most amplification technologies, the main advantages of BB are the faster inference time and the ability to train the model directly on experimental data, the case of bismuth-doped fiber amplifiers is particularly interesting as the BB models are providing physical insight which is not directly available. WB models of Bi-DFAs are not yet well established due to the challenge of fitting the many unknown parameters in standard rate-equation-based models, which is made even more complex by the low concentration of the several types of bismuth active centers~\cite{federico2026numerical}. Nevertheless, beyond low training data efficiency and high computational requirements of BB models, the generalization capability of BB amplifier models is also worsened by the discrete parameter space. For example, in the case of  Raman amplifiers, generalizing beyond one specific fiber type cannot be easily overcome by training with samples from different fiber types, unless all the possible fiber types are considered. This method does not scale favorably in terms of training efficiency. However, the extended training dataset is required by the discretized parameter space as fiber attenuation coefficient, fiber length, Raman efficiency etc. do not span a fully continuous set of values~\cite{deMoura2023JLT}. As an example, typical fiber spans are quantized to specific lengths, which does not allow NN-based models to generalize easily~\cite{deMoura2023JLT}.\\
Architecture-wise, the majority of literature on BB modeling of optical amplifiers relies on feed-forward neural networks (FNNs), either fully trained e.g.,~\cite{Zhou2006,ionescu2020CLEO,minakhmetov2023OFC,deMoura2023JLT} or using extreme learning machine (ELM), e.g. ~\cite{deMoura2020JLT,Donodin2023JEOSRP}. Tandem FNNs, combining a direct FNN model trained together with an inverse FNN model, have also been proposed~\cite{Zibar2019} and convolutional networks have been suggested to capture the spatial-spectral response~\cite{Soltani2021OL}. An example of BB predicted Raman on-off gain is shown in Fig.~\ref{fig:RamanAmpl}(b).

\subsubsection{Directly-modulated lasers}
Laser dynamics exhibit complex behavior that, depending on the driving regime, can encompass bifurcations, multi-attractor states, and chaotic dynamics. Such complexity has motivated the adoption of time-dependent machine learning architectures for their modeling, including reservoir computing~\cite{Amil2019}, long short-term memory (LSTM) networks~\cite{Cheng2022}, Fourier neural operators~\cite{Feng2024}, and transformers~\cite{Zou2025}. However, data-driven modeling of DML has not been investigated as thoroughly as optical amplifiers or laser dynamics, yet. 

In the context of optimizing DML-based communication systems, a model predicting the temporal dynamics that provides faster inference compared to numerically solving (\ref{eq:DML}) with ODE solvers is particularly advantageous. Therefore, BB models of vertical-cavity surface-emitting lasers (VCSELs)~\cite{zhang2025accurate,deligiannidis2026vcsel} and distributed feedback lasers (DFBs)~\cite{fernandez2024PTL} have been reported. The temporal dynamics require context-aware architectures, such as Volterra filters, time-delay neural networks (TDNN), LSTM and transformers. These architectures have been effectively applied to predict the output field of the laser from the input driving current and bias, see Fig.~\ref{fig:NNsketch}(b). However, as the complexity of the dynamics is higher, e.g. compared to optical amplifiers, the model complexity is significantly increased (e.g. from FNN to convolutional-attention transformers, CAT), resulting in increased computational resources required in training but also in inference. Similarly, generalization is worsened by the need to capture complex and sensitive temporal dynamics. In Fig.~\ref{fig:DML}(b)-(d) visual examples of  responses learned by three BB models (TDNN, LSTMs, and convolutional-attention transformers, CATs) are shown through eye diagrams.

\subsubsection{Interferometer meshes}
A NN structure for modeling XT in interferometer meshes is shown in Fig.~\ref{fig:NNsketch}(c)~\cite{Ali_JLT}. This is applied to the $3 \times 3$ MZI mesh to learn the full transfer input-output mapping as a function of voltages. In contrast, for the hexagonal mesh, two BB models, e.g. a simpler linear fit and linear regressor (LR), can map the phases applied to the XT-MZIs to the wavelength shift of the resonances of the MRR cell under test. As for the Raman amplifier, both models are feedforward models since temporal dynamics are not directly relevant, and the models focus on the steady-state response. The physics-agnostic nature of these BB models allows for capturing multiple effects that cannot be easily fitted by WB models. For example, while the involved thermo-optic model in~\cite{marchisio2025comprehensive} effectively improved the prediction's quality compared to purely optical WB models, the overall accuracy is still limited. The WB model relies on design dimensions, thus not accounting for fabrication imperfections and additional electrical and optical XT. A BB model, trained over a sufficiently comprehensive dataset, can capture such effects, and thus improves further the prediction accuracy, as shown in~\cite{Ali_JLT}. Nevertheless, a fully BB model is expected to provide unfavorable scaling of the training dataset with the mesh size. The training dataset consists of randomly applied phases to the MZIs and measured responses. Therefore a comprehensive dataset generally grows super linearly with the number of phases being applied, which, in turn, scales quadratically with the matrix size~\cite{clements_optimal_2016}.

\subsection{Grey-box models}
\label{subsec:GB}
The shortcomings of physics-agnostic BB modeling have triggered a general interest for developing strategies to embed existing physical knowledge into the data-driven models, e.g. replacing only part of the WB model with a BB counterpart or guiding the training of the data-driven model with physical insight. This modeling paradigm combining WB and BB strategies is commonly referred to as grey-box (GB). More specifically, in 2019 the concept of physics-informed NNs was formalized~\cite{raissi_physics-informed_2019}, and since then it has been extensively applied to model photonic systems. 
The main goal of merging data-driven and physics-driven paradigms is to leverage the strengths of each strategy while mitigating their shortcomings. The advantages of GB modeling will be extensively discussed in Section~\ref{sec:Challenges}. Here, we introduce the two main approaches to GB modeling: ML-aided physical modeling and physics-informed data-driven modeling.

\begin{figure}
    \centering
    \includegraphics[width=0.97\linewidth]{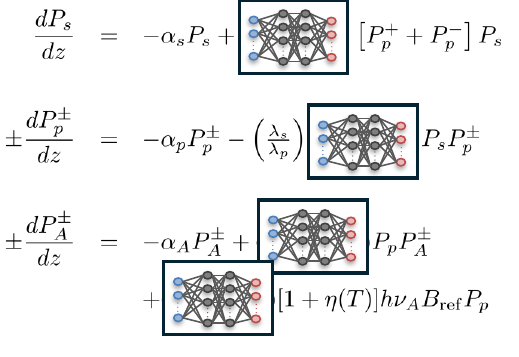}
    \caption{Grey-box modeling - neural-network-aided physical model.}
    \label{fig:GBmodels}
\end{figure}

\subsubsection{Machine-learning-aided physical models}
ML-aided physical modeling refers to only replacing the `challenging' parts of a WB model with data-driven counterparts.
Considering the specific example of Raman amplifiers discussed in Section~\ref{subsec:WB}, the Raman efficiency, $C_R$ makes it challenging to calculate the gradient through the model. Therefore, in~\cite{Yankov2023JLT,marcon2020JLT} it was proposed to replace the integral equation for $C_R$ with its fitted piecewise linear or even NN-based approximation. This proposal is highlighted in Fig.~\ref{fig:GBmodels}.  Choosing this path, the inherent generalization of the system of equations is preserved, while the lack of differentiability is solved. \rev{A detailed comparison of such a modeling approach with conventional white- and black-box models has been reported for a bidirectional Raman amplifier in \cite{yankov2025OFT}, highlighting the trade-offs and advantages of the various methods under that specific target application.}
Similar approaches have also been proposed for lasers and EDFAs. \rev{In~\cite{meinecke2024data}, the microscopic evaluation of the collision term in the rate equations to simulate a semiconductor laser has been replaced with a data-driven model,} reference~\cite{liu2023building} proposed a GB model for an EDFA relying on linearizing and fitting the gain spectrum, while \cite{jones2023spectral} implemented a hybrid model using a NN-based BB model to learn the parameters of a WB model based on~\cite{saleh1990analytical}.

\subsubsection{Physics-informed data-driven models}
Alternatively to ML-aided physical models, the generally available physical insight can guide the data-driven learning (Fig.~\ref{fig:modelingapproaches}(b)), e.g., the model architecture of physics-informed neural networks~\cite{raissi_physics-informed_2019}. Training these GB models is commonly implemented by considering the weighted combination of two loss functions as shown in (\ref{eq:GBloss}). A data-driven loss term $\mathcal{L}_\mathrm{data}$  learns to predict the system's response from data. A physics-driven loss term $\mathcal{L}_\mathrm{physics}$ requires that the model prediction follows the underlying functional dependence described by the WB model, \rev{thus lowering the probability that the prediction will violate physical laws as BB models may do}.  The weight parameter $k$ in (\ref{eq:GBloss}) controls the importance of the data-driven and physics terms during the learning. 

\begin{equation}
\mathcal{L}_{\text{total}} = (1-k)\cdot \mathcal{L}_\mathrm{data}  + k\cdot\mathcal{L}_\mathrm{physics}\,. 
    \label{eq:GBloss}
\end{equation}

Literature on GB modeling for optical amplifiers is also growing, with several interesting proposals being reported recently on Raman amplifiers~\cite{song2024srs,zhang2024CLEO,zhang2025physics}, and fiber-doped amplifiers~\cite{jiang2025physics,wang2025physics}.  As the WB models of optical amplifiers are better established, their use allows for an impressive decrease in the training data required, potentially down to only a few samples~\cite{zhang2024CLEO}.

While for optical amplifiers and DMLs the WB models are already accurate, for interferometer meshes, multiphysics WB models are challenging to define. Therefore, an additional strength of GB modeling is its ability to enhance the training data efficiency and model generalization even when only partial physical information is provided. For example, as discussed in Section~\ref{subsec:Gen}, in~\cite{teofilovic_physics-informed_2025} only the thermal model is considered for the physics loss and it still provides a clear advantage over BB models. More importantly, the parameters of the WB model embedded in $\mathcal{L}_\mathrm{physics}$ can be fitted during the training of the GB model, allowing for a joint fitting and learning procedure, as sketched in Fig.~\ref{fig:modelingapproaches}(b)~\cite{teofilovic_physics-informed_2025}. \rev{We remark that these two approaches are simply common in optical communication and computing. Similar approaches are being used even in terms of equalization of communication systems e.g. through Kolmogorov–Arnold Networks \cite{wang2026physics}. Furthermore, a plethora of methods for combining physics insight with data-driven learning exists, e.g. refer to~\cite{meng2025physics} for a more comprehensive overview of GB models applied to physics. }

\section{Open challenges and potential solutions}
\label{sec:Challenges}
This section moves beyond a general comparison of WB, BB, and GB models to quantitatively examine the main open challenges and trade-offs across these three paradigms, and to highlight promising strategies to address them. A central finding is that combining physical and data-driven training offers clear advantages. The analysis is guided by the goal of developing models for programming, controlling, and optimizing photonic systems, with model complexity, training-data efficiency, generalization, and modularity as the key figures of merit.
\subsection{Model complexity}
\begin{figure}
    \centering
    \includegraphics[width=0.97\linewidth]{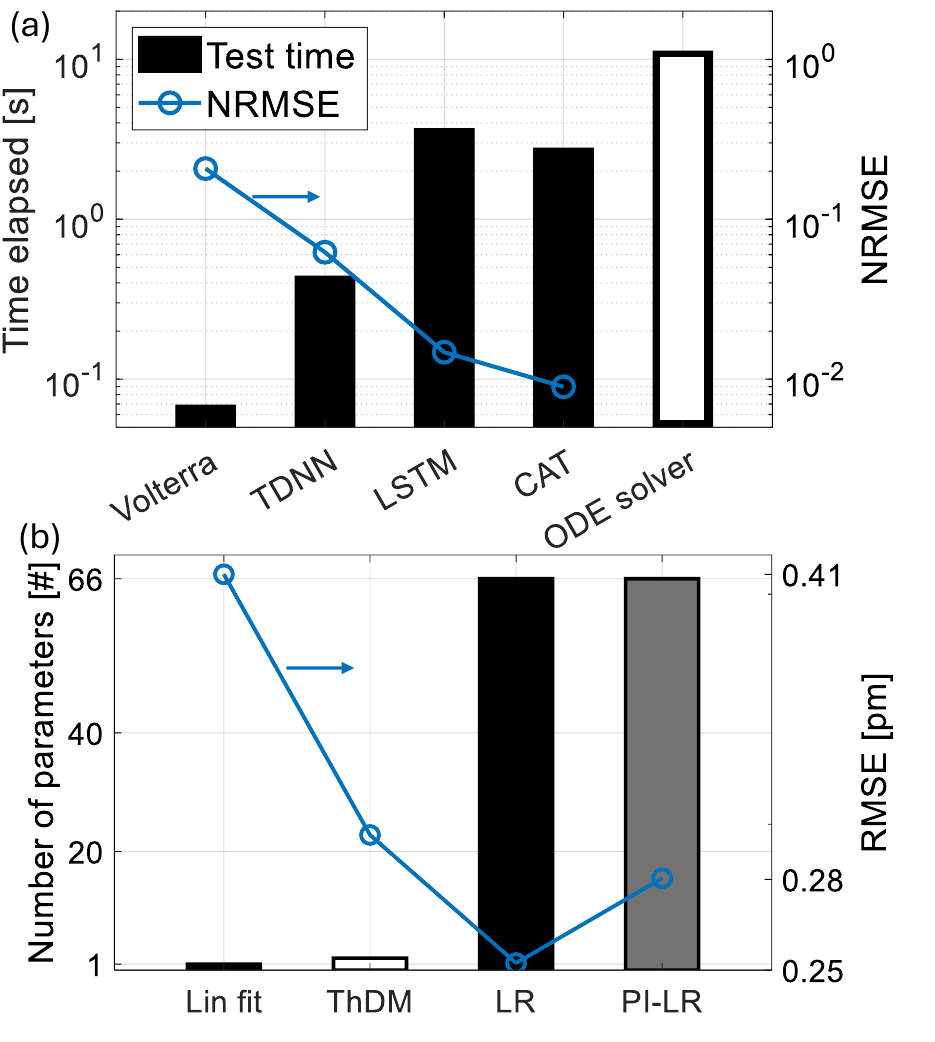}
    \caption{\textbf{Model complexity} - (a) trade-off between inference time and model accuracy for the DML BB (Volterra, TDNN, LSTM and CAT) and WB (ODE solver) models (adapted from~\cite{fernandez2024PTL}) (b) trade-off between model parameters and model accuracy for the hexagonal MZI mesh BB (linear fit and LR), WB (ThDM) and GB (PILR) models.}
    \label{fig:ModelCOmplexity}
\end{figure}

Model complexity is one of the most critical metrics for evaluating model quality. Inference time, i.e. the time required to run the model, becomes especially critical in time-sensitive applications, such as link reconfiguration by optimizing an amplifier gain profile or the signal driving a DML.
Fig.~\ref{fig:ModelCOmplexity} shows the trade-off between model accuracy and model complexity, measured either as inference time for the tested DML BB models (Fig.~\ref{fig:ModelCOmplexity}(a))  or as number of parameters for the MZI mesh models (Fig.~\ref{fig:ModelCOmplexity}(b)). Starting with the DML models, there is a clear correlation between the model performance and the required inference time, here measured over a test dataset of $2^{17}$ data points. Moving from nonlinear Volterra filters or TDNNs to LSTMs or CATs, the RMSE decreases while the inference time grows. Nevertheless, comparing CAT and ODE solver (ground truth), the inference time is still reduced fourfold, at the small price of an NRMSE below 1\%~\cite{fernandez2024PTL}.
A similar trend can be seen for the hexagonal MZI mesh, where the prediction accuracy of the resonance wavelength shift directly scales with the number of fitted parameters in the models. The comparison between WB models (ThDM), BB models (linear fit and linear regressor, LR), and GB model (physics-informed linear regressor, PILR) follows the trend defined by the number of parameters of each model. The GB version of LR (PILR) introduces two additional trainable parameters through $\mathcal{L}_\mathrm{physics}$, which define the physics-based constraint rather than the model itself. As such, the number of parameters governing the mapping between inputs and outputs remains unchanged. Consequently, when testing on data from the same device (self-test), the RMSE of PILR lies between those of ThDM and LR, as expected from a model that combines data-driven learning with a physically motivated constraint.

However, the GB model still provides substantial advantages in terms of generalization, as will be discussed in Section~\ref{subsec:Gen}. 

\subsection{Training data efficiency}

\begin{figure*}
    \centering
    \includegraphics[width=0.97\linewidth]{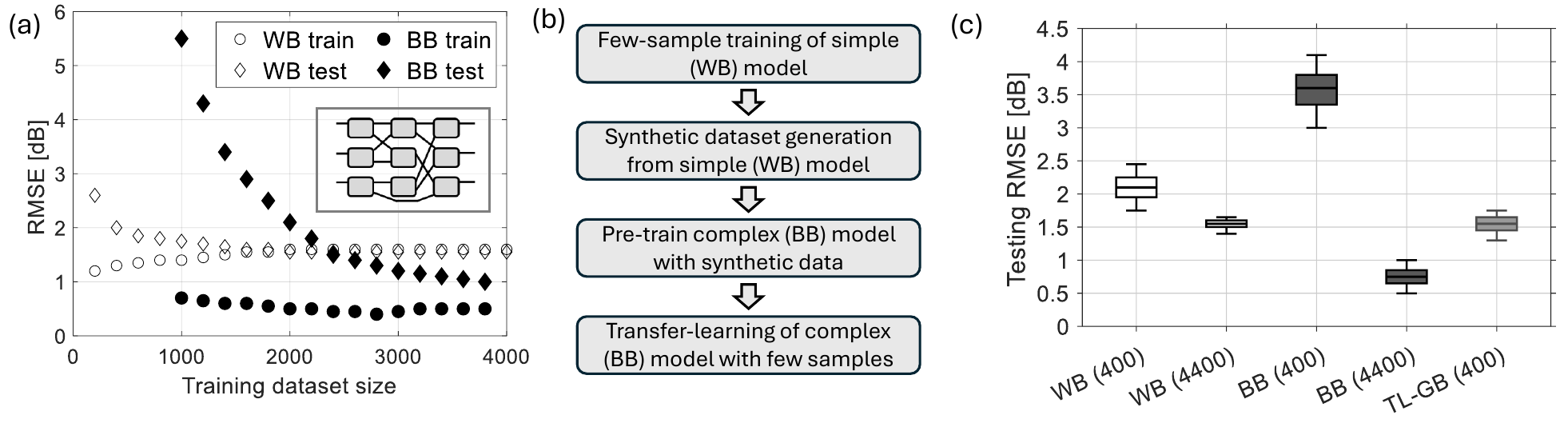}
    \caption{\textbf{Data efficiency} - (a) comparison between training and testing RMSE as a function of training dataset size for WB and BB models of the $3 \times 3$ MZI architecture (inset), adapted from~\cite{cem2023OL}; (b) proposed workflow for enhancing model accuracy under strong training data constraints through transfer learning; (c) performance of WB and BB models compared with TL-GB model, adapted from~\cite{cem2023OL}.}
    \label{fig:DataEff}
\end{figure*}
An inherent consequence of the model complexity, beyond the computing resources required, is the size of the training data needed to allow a model with a larger number of parameters to converge. 

This is directly shown in Fig.~\ref{fig:DataEff}(a) where training and testing RMSE is shown for a $3 \times 3$ rectangular MZI mesh~\cite{cem2023OL}. Simpler WB models, e.g. the model cascading the MZI transfer function in (\ref{eq:MZI}), require 25\% lower training data compared to a more evolved NN-based BB model. Data-efficiency can be directly related to the number of trainable parameters, 81 for the WB model, and over 13500 parameters for the BB model. However, the lower complexity, as discussed in the previous section, directly limits the achievable performance.

In data-constrained settings, such as it is normally the case when it comes to the availability of measurement data, a powerful technique to improve model performance without expanding the training dataset is transfer learning. Transfer learning (TL) was initially applied to Raman amplifiers in~\cite{deMoura2023JLT}. The general workflow is shown in Fig.~\ref{fig:DataEff}(b). Since simpler WB models converge faster, fewer-sample training is sufficient. Once the WB model is trained, a large synthetic data for pre-training of the more complex BB model can be generated. Finally, the pre-trained BB model is fine-tuned, e.g., through transfer learning, with the original few samples used to train the WB model. Note that synthetic dataset generation can be parallelized, unlike sequential iterative optimizations, thus relaxing the computational requirement even for more complex WB models.

In the case of the $3 \times 3$ MZI mesh considered here, the model performance is shown in Fig.~\ref{fig:DataEff}(c). 
When the training dataset is reduced to fewer than 400 samples, i.e. less than 10\% of those required for BB convergence, the WB model already achieves near-converged performance with an RMSE only 30\% above its converged value. In contrast, the BB model exhibits nearly four times worse RMSE than at convergence and twice the error of the WB model under the same conditions~\cite{cem2023OL}. By applying the technique in Fig.~\ref{fig:DataEff}(b), the transfer-learned GB model (TL-GB), so labeled as the training is guided by the synthetic dataset generated through the WB model, improves the BB model performance by more than a factor 2 and with only 0.75 dB worse performance than a BB model trained over the full 4400 training samples.

Additionally, this technique is particularly effective for addressing the related challenge of training data with discretized values, e.g. in the case of Raman amplifiers. As discussed, fiber parameters tend to have discrete values, e.g. length and attenuation. By generating a synthetic dataset, `virtual' fibers with arbitrary parameters can be included in the dataset. The functional dependency between fiber parameters and output gain can then be better interpolated by the BB model, as demonstrated in~\cite{deMoura2023JLT}.

Finally, going a step beyond using the WB model to generate synthetic training data, a fully GB model can be used. 
EDFA GB models~\cite{jones2023spectral,liu2023building}, and Raman amplifier GB models~\cite{song2024srs,zhang2024CLEO} could be trained with only a few samples.

\begin{figure}
    \centering
    \includegraphics[width=0.97\linewidth]{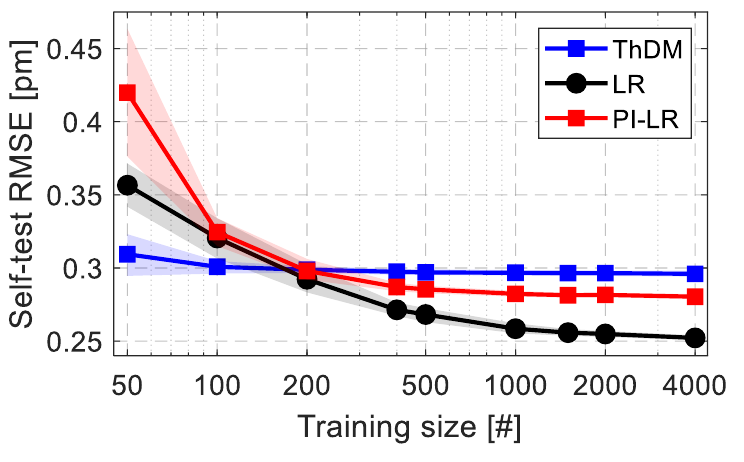}
    \caption{\textbf{Data efficiency} - Self-Testing RMSE as a function of the training dataset size: comparison between ThDM (WB), LR (BB) and PILR (GB) models. Shaded areas refer to the standard deviation over 10 training datasets.}
    \label{fig:DataEff2}
\end{figure}

In Fig.~\ref{fig:DataEff2}, we show the testing RMSE as a function of the training dataset size for WB, BB and GB models of thermal XT in a hexagonal MZI cell. For each training dataset size below 4000, ten distinct training sets were sampled and used to train independent models. Symbols report the mean test error across these ten training runs, while shaded areas indicate one standard deviation.
As discussed for the $3 \times 3$ MZI mesh, the WB model performs best in heavily data-constrained scenarios, e.g., below 200 training data points. Beyond 200 data points, the LR (BB) model provides the lowest error, even compared to its GB alternative (PILR). PILR requires less training data to converge to its best RMSE. However, its best RMSE always lies between the RMSEs of WB and LR for each training dataset size (as discussed in the previous section). Therefore, its higher RMSE makes it a less desirable modeling choice. Nevertheless, as will be discussed in the next section, the better RMSE of the LR model comes from overfitting to the single cell, and leads to poor generalizability, which is instead substantially improved by the PILR models.

\subsection{Generalizability}
\label{subsec:Gen}
Accurate models need to be able to generalize beyond simple overfitting to their training set. In the context of modeling physical devices, generalization should be considered in terms of: (1) capturing the impact on the system response from physical quantities/parameters that can be varied for a single physical device (\emph{`parameter-generalization'}); and (2) predicting the behavior of multiple similar physical devices sharing the same physics, \rev{or even similar parts of the same device} (\emph{`device-generalization'}). 
\emph{Parameter-generalization} can be achieved by extending the training dataset to include data points for different values of each physical parameter. If the hyper-space is sampled sufficiently during training (see discussion in Section~\ref{subsec:BB}), then \emph{parameter-generalization} is achievable. \emph{Device-generalization} can ideally be achieved by providing training data from multiple devices~\cite{DaRos2020ECOC,deMoura2023JLT}. However, for this method the dataset size does not scale favorably with the number of devices. Therefore, alternative paradigms have been sought. As discussed in Section~\ref{subsec:GB}, physical knowledge embedded in the models improves generalization since it provides the model with the embedded functional dependencies common across devices.

\begin{figure}[!ht] 
    \centering
    \includegraphics[width=0.97\linewidth]{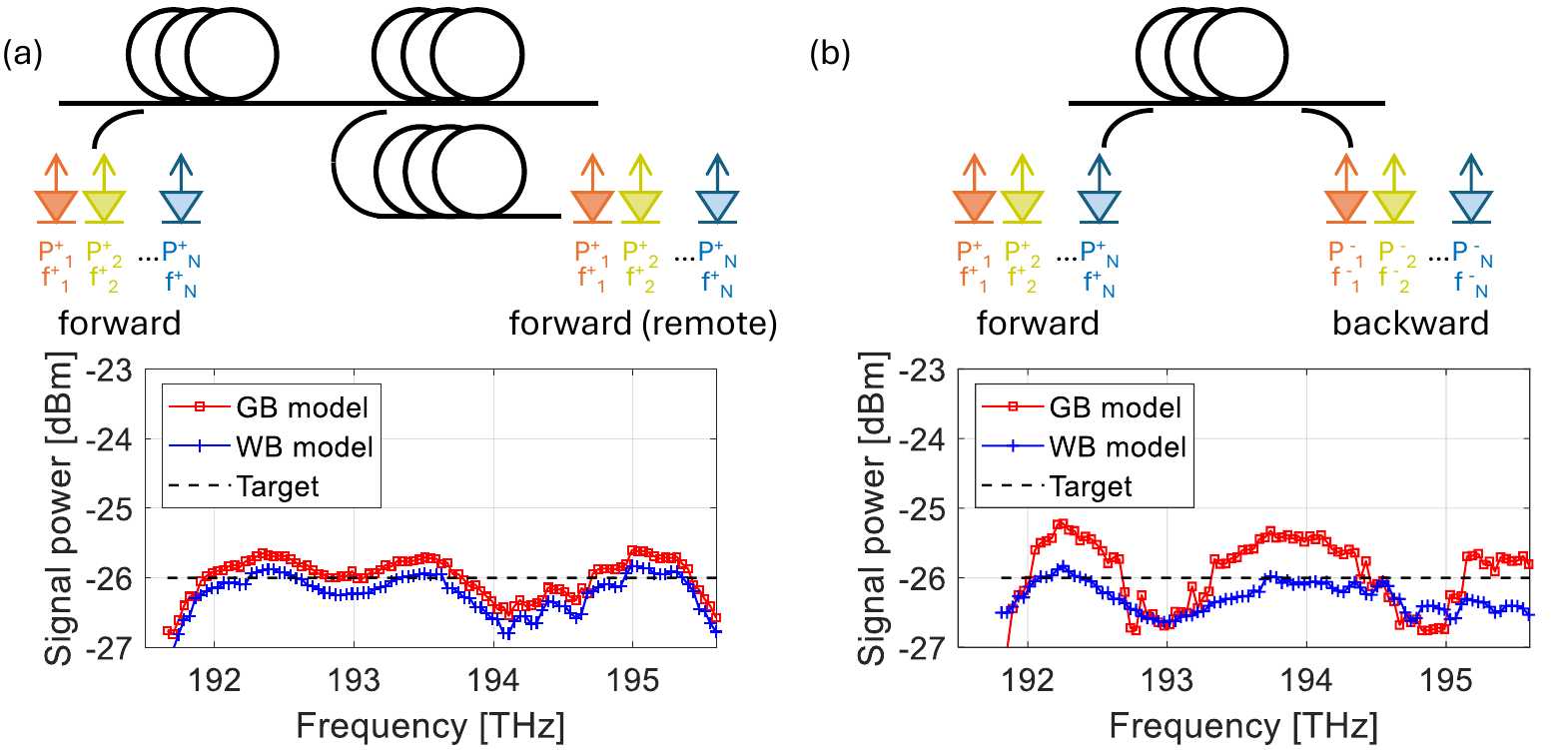}
    \caption{\textbf{Generalizability} - Raman amplifier optimization for a target output power spectral profile for (a) forward pumping across two spans (second span with remote pumping) and (b) bidirectional pumping across a single span. The optimization target is compared with the predicted power from the GB model and the power obtained from a computationally heavier fully WB model, adapted from~\cite{Yankov2023JLT}.}
    \label{fig:GeneralRaman}
\end{figure}

As shown in~\cite{Yankov2023JLT}, ML-aided physical models of Raman amplifiers can generalize across different amplifier types with forward, backward and bidirectional pumping by only varying the input parameters of the model, e.g. the pump powers. This generalizable and differentiable model can then be applied for power optimization of the amplifiers and the results are shown in Fig.~\ref{fig:GeneralRaman}(a) and (b) for a forward and bidirectional amplifier, respectively. The optimized power matches closely the target and more importantly, the model prediction has been independently verified with a fully WB model showing a good agreement~\cite{Yankov2023JLT}.

\begin{figure*}[!htb]
    \centering
    \includegraphics[width=0.97\linewidth]{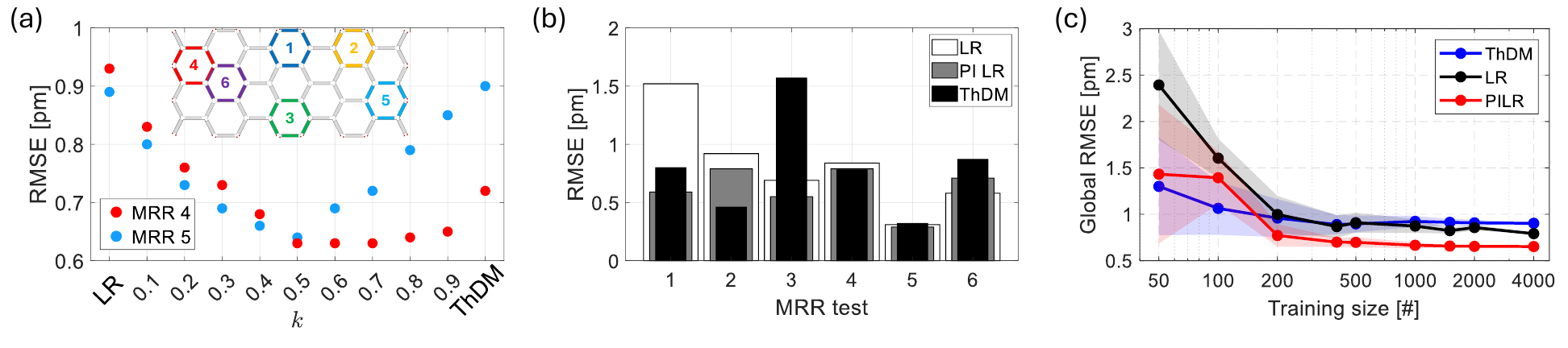}
    \caption{\textbf{Generalizability} - (a) Testing RMSE across all unit cells as a function of the loss-weight parameter for the GB model. (inset) schematic of the programmable PIC highlighting the hexagonal unit cells considered. (b) Comparison of WB, BB and GB based on the RMSE over the test unit cells for the model trained on cell 5. (c) Data-efficiency evaluated through the global RMSE as a function of the training dataset size for the three models. Symbols report mean values and shaded areas one standard deviation. }
    \label{fig:GeneralMVM}
\end{figure*}

Programmable PICs are often realized by cascading identical unit cells, as in interferometer meshes. In this context, generalizable models are of particular interest, as they enable learning the behavior of a single unit cell and subsequently reconstructing the full PIC model through symmetry and translation operations, effectively yielding a modular modeling framework. To analyze modularity, we consider the hexagonal mesh and compare the generalization properties of the same three models analyzed in the previous section: one WB (ThDM), one BB (LR) and one GB (PILR). \rev{In this scenario, we will consider this \emph{cross-cell generalization} as an example of device generalization, by considering the single device as the hexagonal cell.} Owing to the discrete and symmetric topology of the hexagonal mesh, the distances between XT-MZIs and a given mesh hexagonal cell form a finite set of distinct values. The number of distinct values depends on which cell is programmed as the MRR. Consequently, ThDM can be represented as a look-up table with one coefficient per distance. LR and PILR, on the contrary, assign one coefficient per XT-MZI. To enable comparison and generalization across the mesh, coefficients corresponding to the same distance are averaged, yielding a single distance-dependent coefficient. Models trained on cells 4 and 5 can thus be reused across the entire mesh, as these cells see all distances present in the PIC. Accordingly, for each modeling type, two models are trained using data from cells labeled as '4' and '5' in Fig.~\ref{fig:GeneralMVM}(a, inset). The six models are then tested across all six cells, including testing on unseen data from the same cell and on data from other cells across the PIC as explained above. 

First, the GB models are optimized by considering the balance between physics-loss and data-driven loss. The resulting testing RMSE values evaluated using test data from all 6 cells (global RMSE) are shown in Fig.~\ref{fig:GeneralMVM}(a). While the models trained on cell 4 and 5 show slightly different performance, in both cases a weighting factor $k=0.5$, i.e. an equal balance between physics-loss and data-driven loss, provides the best \emph{device-generalization}. In Fig.~\ref{fig:GeneralMVM}(b), the comparison between WB, BB and GB models is reported for the models trained on cell 5. For the GB model, the optimized $k=0.5$ is used. As can be seen, the BB model is generally the best performing when tested on cell 5, i.e. self-testing, with the GB model a close second and the WB model with slightly worse performance. However, the GB model provides the best \emph{device-generalization}, with the WB model still showing decent RMSE when tested over different devices, while the BB model is relatively accurate for some cells but shows a drastic increase in RMSE for others. \rev{The generalization of the proposed approach of a joint optimization of $k$ across multiple cells is therefore validated, showing the accuracy gain on each of the single cells.}

The averaged testing RMSEs across the 6 cells are listed in Table~\ref{tb:GeneralMVM}, confirming the trends from Fig.~\ref{fig:GeneralMVM}(b).
\begin{table}[!htb]
    \centering
    \begin{tabular}{l|c|c}
       model  &  train RMSE [pm] & test RMSE [pm] \\\hline\hline
       WB  & 0.31 & 0.90 \\\hline
       BB  & 0.28 & 0.90 \\\hline
       GB  & 0.27 & 0.63\\
    \end{tabular}
    \caption{\textbf{Generalizability} - Train and global (test) RMSE values for WB, BB and GB models. Each model is trained only with training data from cell 5, while the global RMSE is averaged across the test data from all 6 cells.}
    \label{tb:GeneralMVM}
\end{table}

Finally, in Fig.~\ref{fig:GeneralMVM}(c), the same three models are compared in terms of the global RMSE as a function of the training dataset size. This analysis extends the analysis of Fig.~\ref{fig:DataEff2}, focusing on \emph{device-generalization}, in this context considered as cross-cell generalization. As for Fig.~\ref{fig:DataEff2}, symbols report the mean values over 10 training runs, while the shaded areas show one standard deviation. Similarly to the self-testing RMSE of Fig.~\ref{fig:DataEff2}, the ThDM performs best for highly data-constrained scenarios, e.g., below 200 training data points. However, above 200 training points, the PILR always outperforms its physics-agnostic counterpart. This analysis confirms the better performance of LR over PILR seen for the self-testing RMSE was due to overfitting to the single cell. 

\subsection{Modular integration}
Once \emph{device-generalizable} models are available, the final goal is to build a full system model by modularly combining models. 

\begin{figure}
    \centering
    \includegraphics[width=0.97\linewidth]{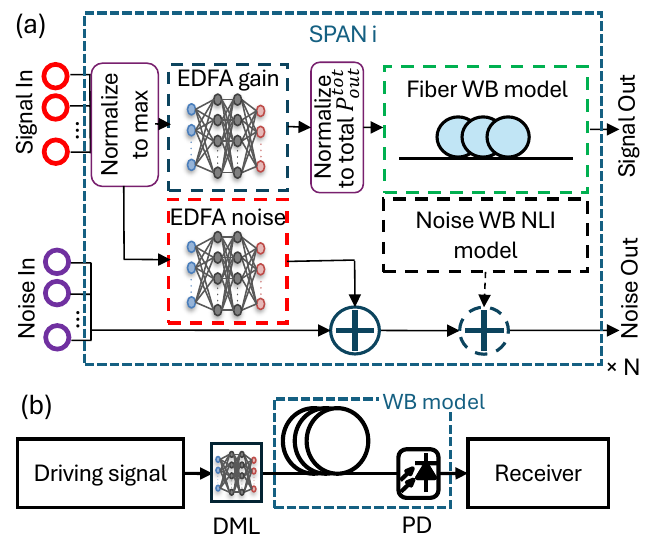}
    \caption{\textbf{Modularity} - Cascadability of (a) BB amplifier models with WB fiber models for optimizing coherent transmission systems (adapted from~\cite{yankov2021snr}, NLI: nonlinear interference) and (b) BB DML and WB fiber models for optimizing short-reach systems.}
    \label{fig:ModularTx}
\end{figure}

Implementing differentiable models of the various components is a critical step towards end-to-end system optimization, e.g. to increasing transmission throughput~\cite{yankov2021snr} or decreasing complexity~\cite{nielsen2025JLT}. Interconnecting these differentiable models modularly is another. 
Several models have been proposed to model the propagation through optical fibers~\cite{hasegawa1973transmission,poggiolini2013gn}. They usually provide a trade-off between accuracy and computational efficiency, e.g. Gaussian noise (GN) model~\cite{poggiolini2013gn} vs. the split-step Fourier method, and can all be implemented to be differentiable~\cite{gaiarin2020end}. Therefore, interconnecting accurate and generalizable optical amplifier models with fiber propagation models allows for full system optimization. An example is provided in Fig.~\ref{fig:ModularTx}(a) where a multispan system is modeled by sequentially cascading (1) a generalizable EDFA model predicting output PSD and noise figure for a given input PSD and pump conditions and (2) an analytical fiber model~\cite{ferrari2020gnpy}. The offline optimization using such a model has been validated experimentally showing excellent agreement~\cite{yankov2021snr}

Similarly, for short-reach interconnects, a DML model could be cascaded with the above-mentioned fiber transmission model as shown in Fig.~\ref{fig:ModularTx}(b). The key challenge in achieving this goal is to build a fully independent DML model, e.g., not including the receiver photodetector as in~\cite{fernandez2024PTL,srinivasan2023JLT}. Such a task requires training the model on both the photon number (intensity) and the phase (signal phase), the latter being more experimentally involved to measure. No modular DML model has yet been reported, to the best of our knowledge.

\begin{figure*}
    \centering
    \includegraphics[width=0.97\linewidth]{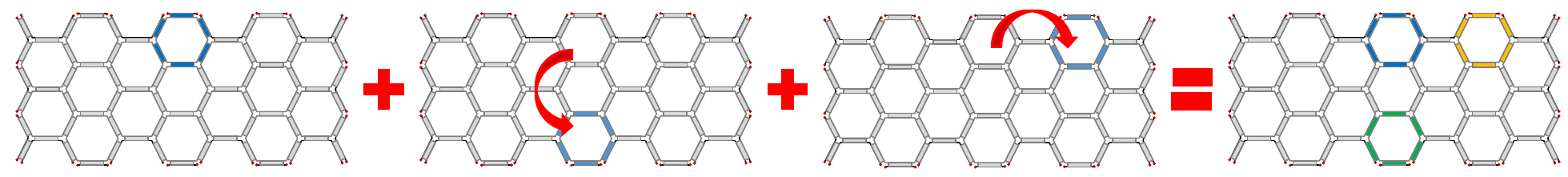}
    \caption{\textbf{Modularity} - Conceptual modularity of crosstalk in multi-cell MZI meshes: translation of a single-cell generalizable model and linear combination of the translated models provides the model of the full PIC.}
    \label{fig:ModularMVM}
\end{figure*}

Finally, programmable PICs also require the ability to accurately program the different areas of the PIC to implement the desired transfer functions. As for the cascadability of the generalizable amplifier models, a generalizable model of a hexagonal cell of an MZI mesh can be translated to model the XT affecting other cells, see Fig.~\ref{ModularMVM}. More importantly, thermal XT effects are expected to sum up linearly. Therefore, XT prediction and compensation can be achieved by solving the system of coupled linear equations, here discussed for the case of programming $N$ MRRs onto the same mesh. \rev{The following derivation considers a uniform phase correction applied to the six MZIs forming each MRR, assuming that these corrections can be assigned independently to the different emulated MRRs.} For such a mesh, the compensated phase vector of MRR$_i$ is given by:
    
\begin{equation}
\Phi_i = \Phi_i^0 - c_i \Delta\lambda_i \mathbf{1}\,,
\label{eq:phi_comp_general}
\end{equation}

\noindent where $\Phi_i^0$ denotes the phase vector required to implement MRR$_i$ in the absence of thermal XT, $\mathbf{1}$ is a vector of ones of length 6 corresponding to the 6 MZIs forming a MRR, and  $c_i = \frac{2\pi}{6 \cdot FSR}$. Assuming the linearity of XT, the total wavelength shift experienced by MRR$_i$ can be written as

\begin{equation}
\Delta\lambda_i = \sum_{\substack{j=1 \ j\neq i}}^{N} \Delta\lambda_{i,j}\,,
\end{equation}

\noindent where $\Delta\lambda_{i,j}$ is the contribution induced by MRR $j$. Each contribution is modeled as

\begin{equation}
\Delta\lambda_{i,j} = \mathbf{a}_{i,j}\Phi_j\,,
\end{equation}

\noindent \{$\mathbf{a}_{i,j}$\} contains the contributions of the MZI forming MRR $j$ to the wavelength shift of the resonances of MRR $i$. For the ThDM model, the elements of ($\mathbf{a}_{i,j}$) are given by applying (\ref{eq:ThDM_temp}), whereas for LR-type models they correspond directly to the learned weights. Substituting (\ref{eq:phi_comp_general}) into the previous expression yields

\begin{equation}
\Delta\lambda_{i,j}
=
b_{i,j}
-
c_j A_{i,j}\Delta\lambda_j\,,
\end{equation}

\noindent where

\[
b_{i,j} = \mathbf{a}_{i,j}\Phi_j^0\,,
\qquad
A_{i,j} = \mathbf{a}_{i,j}\mathbf{1}\,.
\]

\noindent Summing over all $j\neq i$ gives

\begin{equation}
\Delta\lambda_i
+
\sum_{\substack{j=1 \ j\neq i}}^{N}
c_j A_{i,j}\Delta\lambda_j
=
\sum_{\substack{j=1 \ j\neq i}}^{N}
b_{i,j}\,.
\label{eq:general_system}
\end{equation}

\noindent Equation~(\ref{eq:general_system}) defines a coupled linear system that can be expressed compactly as

\begin{equation}
\mathbf{M}\Delta\boldsymbol{\lambda} = \mathbf{b}\,,
\end{equation}

\begin{figure}
    \centering
    \includegraphics[width=0.97\linewidth]{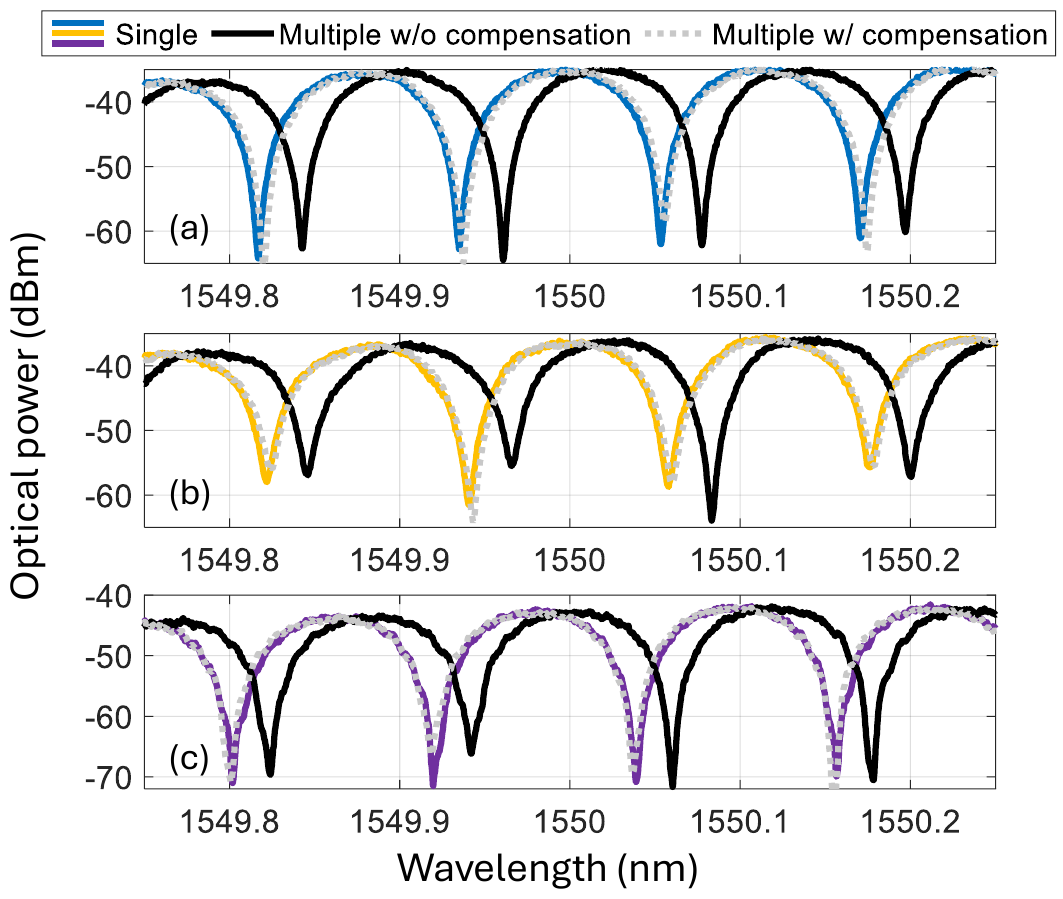}
    \caption{\textbf{Modularity} - Modular Thermal XT compensation - Spectral response of MRR$_i$ when programmed alone compared with its response when the full mesh (MRR$_1$, MRR$_2$, MRR$_6$) is programmed simultaneously, with and without XT compensation, for (a) $i=1$, (b) $i=2$, and (c) $i=6$. Wavelength shifts are predicted using the PILR model trained on MRR$_5$.}
    \label{fig:ModularMRRs}
\end{figure}

\noindent where \rev{$\Delta\boldsymbol{\lambda}=[\Delta\lambda_1, ..., \Delta\lambda_N]^\mathrm{T}$ is an $N$-dimensional vector whose entries correspond to the total wavelength shifts experienced by the $N$ programmed MRRs, and $\mathbf{M}$ is an $N \times N$ matrix with elements}:

\[
M_{i,j} =
\begin{cases}
    1, & i=j\,, \\
    c_j A_{i,j}, & i\neq j\,.
\end{cases}
\]

\noindent and

\[
b_i =
\sum_{\substack{j=1 \ j\neq i}}^{N}
b_{i,j}\,.
\]

\noindent The compensated phase vectors are then obtained by solving

\begin{equation}
\Delta\boldsymbol{\lambda}
=
\mathbf{M}^{-1}\mathbf{b}\,,
\label{eq:MultiXT}
\end{equation}

\noindent and substituting the resulting wavelength shifts into (\ref{eq:phi_comp_general}), thereby compensating for the thermal XT and obtaining the phases to apply to simultaneously implement all $N$ MRRs accurately in the presence of thermal XT. \rev{As a preliminary demonstration of the efficacy of the method, we simultaneously program MRR$_1$, MRR$_2$, and MRR$_6$ on the same mesh, with additional XT-MZIs turned on (numbering follows the inset in Fig. 11(a)). The goal is to predict the resulting thermal XT using (\ref{eq:MultiXT}) and compensate for it so that each MRR's spectral response remains consistent whether programmed alone or as part of the full mesh.}

\rev{Fig.~\ref{fig:ModularMRRs} compares the spectral response of each MRR in isolation against its response when the full mesh is programmed, with and without compensation, for i = 1 (a), i = 2 (b), and i = 6 (c). In all cases, the wavelength shift is predicted using the PILR model trained on MRR5.}

\section{Conclusion}
\label{sec:Conclusions}
Accurate, rapid, generalizable, and differentiable models of photonic components and subsystems are essential for the effective control, programming, and optimization of increasingly complex photonic integrated circuits and optical systems. This work presents a comparative analysis of three prominent modeling paradigms: white-box physics-based, black-box data-driven, and grey-box physics-informed data-driven approaches. By evaluating these paradigms across three distinct use cases - optical amplifiers, directly modulated lasers, and interferometer meshes - we examined their respective advantages and limitations. The quantitative comparison highlights that\rev{, at least within the scope of these use cases,} grey-box models offer significant advantages, substantially enhancing generalizability and modularity while preserving training data efficiency and limiting model complexity. Accordingly, the integration of physical insight into data-driven models provides a robust paradigm for the efficient control, programming, and optimization of complex optical systems.



\ifCLASSOPTIONcaptionsoff
  \newpage
\fi

%
\bibliographystyle{IEEEtran}
\bibliography{refGen,refAmp,refDML,refMZI}

\end{document}